\documentclass{emulateapj}

\shorttitle{First Supernova Explosions}

\shortauthors{Greif et al.}

\begin{document}

\title{The First Supernova Explosions:\\Energetics, Feedback, and Chemical Enrichment}

\author{Thomas H. Greif\altaffilmark{1,2,3}, Jarrett L. Johnson\altaffilmark{2}, Volker Bromm\altaffilmark{2}, and Ralf S. Klessen\altaffilmark{1}}

\altaffiltext{1}{Institut f\"{u}r theoretische Astrophysik, Albert-Ueberle Strasse 2, 69120 Heidelberg, Germany; tgreif@ita.uni-heidelberg.de, ralf@ita.uni-heidelberg.de}
\altaffiltext{2}{Department of Astronomy, University of Texas, Austin, TX 78712; jljohnson@astro.as.utexas.edu, vbromm@astro.as.utexas.edu}
\altaffiltext{3}{Fellow of the International Max Planck Research School for Astronomy and Cosmic Physics at the University of Heidelberg}

\begin{abstract}
We perform three-dimensional smoothed particle hydrodynamics simulations in a realistic cosmological setting to investigate the expansion, feedback, and chemical enrichment properties of a $200~M_{\odot}$ pair-instability supernova in the high-redshift universe. We find that the SN remnant propagates for a Hubble time at $z\simeq 20$ to a final mass-weighted mean shock radius of $2.5~\rm{kpc}$ (proper), roughly half the size of the H~{\sc ii} region, and in this process sweeps up a total gas mass of $2.5\times 10^{5}~M_{\odot}$. The morphology of the shock becomes highly anisotropic once it leaves the host halo and encounters filaments and neighboring minihalos, while the bulk of the shock propagates into the voids of the intergalactic medium. The SN entirely disrupts the host halo and terminates further star formation for at least $200~\rm{Myr}$, while in our specific case it exerts positive mechanical feedback on neighboring minihalos by shock-compressing their cores. In contrast, we do not observe secondary star formation in the dense shell via gravitational fragmentation, due to the previous photoheating by the progenitor star. We find that cooling by metal lines is unimportant for the entire evolution of the SN remnant, while the metal-enriched, interior bubble expands adiabatically into the cavities created by the shock, and ultimately into the voids with a maximum extent similar to the final mass-weighted mean shock radius. Finally, we conclude that dark matter halos of at least $M_{\rm{vir}}\ga 10^{8}~M_{\odot}$ must be assembled to recollect all components of the swept-up gas.
\end{abstract}


\section{Introduction}
One of the main goals in modern cosmology is to understand the formation of the first stars at the end of the cosmic dark ages and how they transformed the homogeneous primordial universe into a state of ever increasing complexity \citep[e.g.,][]{bl01,miralda03,bl04a,cf05,glover05}. These so-called Population~III (Pop~III) stars, predicted to be very massive, with $M_{*}\ga 100~M_{\odot}$, and forming in DM ``minihalos'' with virial masses $M_{\rm{vir}}\ga 10^{6}~M_{\odot}$ \citep[e.g.,][]{bcl99,bcl02,nu01,abn02,yoshida06,gao07,on07}, were responsible for the initial metal enrichment of the intergalactic medium (IGM). The first supernova (SN) explosions rapidly dispersed the heavy elements that were produced during the brief lifetime of a Pop~III star into the IGM, thus beginning the long nucleosynthetic build-up from a pure H/He universe to one with ubiquitous metal enrichment \citep[e.g.,][]{fps00,mfr01,mfm02,sfm02,tsd02,byh03,fl03,fl05,wv03,daigne04,daigne06,nop04,ybh04,ky05}.

The first SNe exerted important chemical and mechanical feedback effects on the early universe \citep[e.g.,][]{cf05}. First, the character of star formation changed from an early, high-mass--dominated (Pop~III) mode to a more normal, lower mass (Pop~II) mode, once a critical level of enrichment had been reached, the so-called critical metallicity, $Z_{\rm{crit}}\la 10^{-3.5}Z_{\odot}$ \citep[e.g.,][]{omukai00,bromm01,bl03a,schneider03,schneider06,fjb07,ss07}. It is then crucially important to understand the topology of early metal enrichment and when a certain region in the universe becomes supercritical \citep[e.g.,][]{schneider02,mbh03,ssf03,ro04,fl05,mc05,gb06,venkatesan06}. Second, the SN blast waves mechanically impacted the halos that hosted Pop~III stars by heating and subsequently evacuating the dense gas inside them. Such a negative feedback effect, limiting the capacity for further star formation, has previously been considered for low-mass galaxies with correspondingly shallow potential wells \citep[e.g.,][]{larson74,ds86,mf99} but also for halos in the vicinity of a SN progenitor, where the net effect is still uncertain \citep[e.g.,][]{cr07}. A qualitatively different mechanical feedback effect has recently been suggested to occur in the dense, post-shock regions of the energetic blast wave \citep{mbh03,sfs04,machida05} where secondary star formation might be induced by the onset of gravitational instabilities, thus giving rise to positive feedback.

The investigation of energetic explosions in the high-redshift universe has a long and venerable history. Most of this early work has focused on the cooling and fragmentation of astrophysical blastwaves \citep[e.g.,][]{bertschinger85,vob85,wandel85} and the possibility of self-propagating galaxy formation \citep[e.g.,][]{ikeuchi81,oc81,cba84}. Such explosive galaxy formation models typically implied substantial distortions to the cosmic microwave background blackbody spectrum and were thus excluded by the {\it Cosmic Background Explorer} \citep[{\it COBE};][]{fixsen96}. Many of the physical effects discussed in those seminal papers, however, remain relevant for our current studies.

What kind of SN explosion is expected to end the life of a massive Pop~III star? According to the precise progenitor mass, the SN could be either of the conventional core-collapse type (for masses $\la 40~M_{\odot}$) or it could be a pair-instability supernova (PISN) for masses in the range $\sim 140$--$260~M_{\odot}$ \citep{hw02,heger03}. In this paper, we specifically consider a PISN with a progenitor mass of $200~M_{\odot}$ and an explosion energy of $E_{\rm{sn}}=10^{52}~\rm{ergs}$. A PISN is predicted to completely disrupt the star, so that all the heavy elements produced will be released into the IGM \citep[e.g.,][]{fwh01,hw02}, resulting in extremely high metal yields. However, our simulation would also approximately describe a hypernova explosion, in which a rapidly rotating, massive star undergoes core collapse \citep[e.g.,][]{un02,tun07}, as far as the energetics and overall dynamics of the blast wave expansion are concerned.

Theoretically, one would expect that a PISN could only have occurred in the early universe, where Pop~III stars were born massive and mass loss might have been negligible \citep{bhw01,kudritzki02}. The recent discovery of the extremely luminous, relatively nearby SN~2006gy, which has tentatively been interpreted as a PISN (Smith et al. 2007; but see Ofek et al. 2007), might defy this theoretical expectation. Having such a nearby example of a PISN, in effect, would provide us with an ``existence proof,'' opening up the exciting possibility of detecting the first SNe with the upcoming {\it James Webb Space Telescope} ({\it JWST}), which will be sensitive enough to observe a single PISN out to $z\sim 15$ \citep[e.g.,][]{mbh03,scannapieco05,wl05,wa05,gardner06}.

In this paper, we carry out cosmological simulations of the first SN explosions, using the smoothed particle hydrodynamics code GADGET in its entropy-conserving formulation \citep[ver. 1.1;][]{syw01}. To address the questions outlined above, it is crucial to investigate the evolution of the SN remnant in three dimensions, starting from realistic initial conditions and including all the relevant physics. In particular, one needs to implement a comprehensive model for the cooling and the chemical evolution of the gas, as well as an efficient algorithm to treat the radiative transfer around the Pop~III progenitor star \citep[see][]{jgb07}. Our simulations greatly improve on earlier work \citep[e.g.,][]{byh03} in that we calculate the preexplosion situation with much greater realism, using a ray-tracing method to determine the structure and extent of the H~{\sc ii} region. In addition, we follow the evolution to much later times, allowing us to reach the point where the blast wave finally stalls and effectively dissolves into the general IGM. We are thus able to analyze the expansion and cooling properties of the SN remnant in great detail, allowing us to draw robust conclusions on the temporal behavior of the shock, its morphology, the final shock radius, and the total swept-up mass.

The structure of our work is as follows. In \S2, we describe the SN progenitor and our method of initializing the SN explosion, followed by a test simulation to verify the accuracy of our results. Subsequently, we discuss the cosmological setup of the main simulation and our treatment of the H~{\sc ii} region. We then concentrate on the evolution of the SN remnant, finding a simple analytic model that summarizes its expansion properties (\S3). In \S4, we investigate the mechanical feedback of the blast wave on neighboring minihalos and elaborate on the prospect of triggering gravitational fragmentation in the dense shell. In \S5, we discuss the relevance of metal cooling, the coarse-grain dispersal of metals, and the general mixing efficiency. Finally, in \S6 we summarize our results and discuss important cosmological implications. For consistency, all distances are physical (proper) unless noted otherwise.

\section{Numerical Methodology}

The treatment of SN explosions in smoothed particle hydrodynamics (SPH) simulations is a demanding problem, primarily due to the strong discontinuities arising at the shock front. One must set up the initial conditions with great care, and it is necessary to enforce strict timescale constraints in calculating the subsequent evolution. In light of these challenges, we describe the SN progenitor and our method of initializing the SN explosion, followed by a performance test to verify the correct behavior of the shock.

\subsection{The Supernova Progenitor}
Although the initial mass function (IMF) of Pop~III is poorly constrained, numerical simulations have indicated that stars forming in primordial halos typically attain $100~M_{\odot}$ by efficient accretion and might even become as massive as $500~M_{\odot}$ \citep[e.g.,][]{op03,bl04b,on07}. \citet{hw02} have discussed the fate of such massive stars and found that in the range $140$--$260~M_{\odot}$ a PISN disrupts the entire progenitor, with explosion energies ranging from $10^{51}$ to $10^{53}~\rm{ergs}$ and yields up to $y\simeq 0.5$.

In the present work, we aim to investigate the most representative case, assuming a stellar mass of $M_{*}=200~M_{\odot}$. We conservatively adopt an explosion energy of $E_{\rm{sn}}=10^{52}~\rm{ergs}$ and a yield of $y=0.1$, to include the possibility of a less top-heavy IMF and a hypernova explosion, but we caution that such models generally predict energy input via bipolar jets, which may invalidate the assumption of a spherically symmetric blast wave. In light of the uncertain progenitor, we note that the dynamics of the SN remnant are governed mainly by the explosion energy, together with the IGM density distribution, and that the composition and metal content of the stellar ejecta become important only at very late times when the enriched gas recollapses (see \S5.1).

\subsection{Energy Injection}
During the early evolution of the SN remnant, i.e., typically less than $10^{4}~\rm{yr}$ after the explosion, the stellar ejecta are confined to a thin shell propagating at constant velocity into the surrounding medium, while secondary shocks quickly bring the interior to a uniform temperature \citep[e.g.,][]{gull73}. This stage in the evolution of the SN remnant is termed the free-expansion (FE) phase, and it lasts until the swept-up mass becomes comparable to the ejecta mass.

To reproduce these well-known initial conditions, we inject the kinetic energy of the SN as thermal energy into the $N_{\rm{sn}}=M_{*}/m_{\rm{sph}}$ innermost particles surrounding the center of the box, where $m_{\rm{sph}}\simeq 5~M_{\odot}$ corresponds to the SPH particle mass. This increases the thermal energy of all particles inside $r_{\rm{fe}}$ to $E_{\rm{th}}=E_{\rm{sn}}/N_{\rm{sn}}$, and the resulting temperature gradient acts as a piston on the surrounding gas. The ensuing shock accelerates the material in the vicinity of $r_{\rm{fe}}$ and creates a thin shell of highly supersonic material, initiating the Sedov-Taylor (ST) phase of the SN remnant. In addition, this method allows us to track the dispersal of metals, as each particle inside $r_{\rm{fe}}$ represents the original stellar content. Although the so-obtained resolution is crude, we can nevertheless quantify the coarse-grain chemical enrichment properties of the SN.

\subsection{Test Simulation}
To test our method of setting up the initial conditions, and also verify that the code reliably calculates the shock propagation, we compare the radial profiles of a test simulation with the well-known ST solution. For this purpose we place the SN according to the above prescriptions in a noncosmological box of length $1~\rm{kpc}$ with $200^{3}$ gas particles and switch off gravity, chemistry, and cooling. The particles are distributed randomly, so that the density fluctuates slightly around the mean of $n_{\rm{H}}=0.5~\rm{cm}^{-3}$, while the temperature is set to $2\times 10^4~\rm{K}$. This choice reflects the situation in the vicinity of a $200~M_{\odot}$ Pop~III star after the surrounding medium has been photoheated \citep[e.g.,][]{kitayama04,wan04,abs06,jb07}, but it also reproduces the initial conditions of the main simulation (see Fig. 8).

Once we initialize the simulation, the shock propagates into the surrounding medium according to the ST solution. As the duration and time-dependent properties of the gas in this phase are crucial for the late-time behavior of the SN remnant, we briefly review the relevant details.

\subsubsection{ST Solution}
For a strong shock in an adiabatic gas, the Rankine-Hugoniot jump conditions imply for the density, velocity, pressure, and temperature directly behind the shock
\begin{eqnarray}
\rho_{\rm{sh}}&=&4\rho\\
v_{\rm{sh}}&=&\frac{3}{4}\dot{r}_{\rm{sh}}\\
P_{\rm{sh}}&=&\frac{3}{4}\rho \dot{r}_{\rm{sh}}^{2}\\
T_{\rm{sh}}&=&\frac{3}{16}\frac{\mu m_{\rm{H}}}{ k_{\rm{B}}}\dot{r}_{\rm{sh}}^{2}\mbox{\ ,}
\end{eqnarray}
where $\rho$ is the density of the surrounding medium, $\mu$ its mean molecular weight, $m_{\rm{H}}$ the mass of the hydrogen atom, and $k_{\rm{B}}$ Boltzmann's constant. Solving the continuity, Euler, and energy equations, one can relate these quantities at {\it any point} behind the shock to those {\it directly} behind the shock as a function of position. Figure 1 shows the resulting self-similar, time-independent character of the ST solution, implying that most of the swept-up mass piles up just behind the shock and forms a dense shell, while the interior regions remain at high temperatures.

Completing the ST solution, the absolute position of the shock as a function of time is given by:
\begin{equation}
r_{\rm{sh}}=\beta\left(\frac{E_{\rm{sn}}t_{\rm{sh}}^{2}}{\rho}\right)^{1/5}\mbox{\ ,}
\end{equation}
where $\beta=1.17$ for an adiabatic gas. Thus, given the SN explosion energy and the density of the surrounding medium, all quantities of relevance around the progenitor are well-determined.

\begin{figure}
\begin{center}
\includegraphics[width=8.0cm,height=5.0cm]{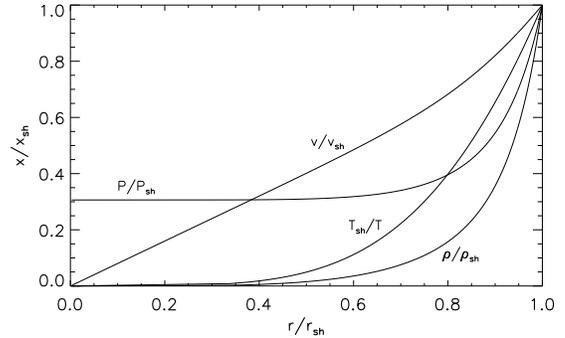}
\caption{ST solution: density, velocity, pressure, and inverse temperature as a function of position relative to the shock. Most of the mass piles up just behind the shock and forms a dense shell, while the isobaric interior regions remain at high temperatures.}
\end{center}
\end{figure}

\subsubsection{Test Results}
 Using the above results, we investigate whether the test simulation has successfully reproduced the ST solution. Figure 2 compares the profiles of the test simulation after $1~\rm{Myr}$, when enough time has passed to enable a relaxation to the asymptotic solution, to the analytic, but dimensionalized, ST profiles shown in Figure 1. Apart from minor deviations caused by initial density fluctuations and higher order shocks, in particular in regard to the velocity distribution, we find that the simulation and the ST solution are in good agreement. The slight offset at the position of the shock is inevitable in light of kernel smoothing.

Being assured that we can reliably treat the propagation of shocks, we turn to the main focus of this work and describe the setup for a SN explosion in the high-redshift universe.

\begin{figure}
\begin{center}
\includegraphics[width=8.0cm,height=7.5cm]{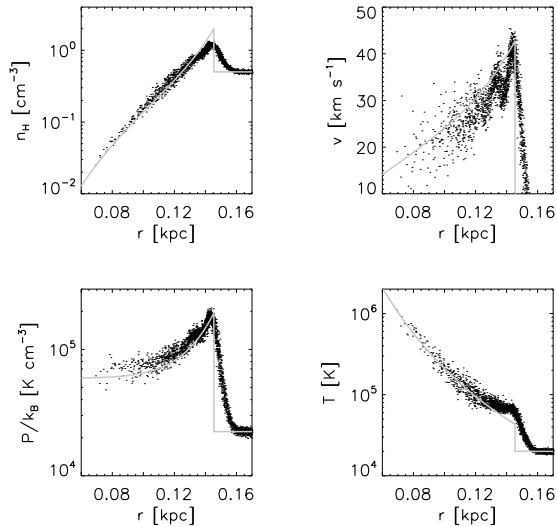}
\caption{Density, velocity, pressure, and temperature of the shocked gas after $1~\rm{Myr}$. Black dots represent the test simulation, while the gray lines show the dimensionalized ST solution. Apart from deviations caused by higher order shocks and kernel smoothing, the simulation reproduces the analytic profiles relatively well.}
\end{center}
\end{figure}

\subsection{Main Simulation}
\subsubsection{Initial Setup}
To investigate the long-term evolution of the SN remnant, but also to correctly calculate the structure and extent of the preceding build-up of the H~{\sc ii} region around the progenitor, we choose a cosmological box of length $150/h~\rm{kpc}$ (comoving), with $200^{3}$ particles per species (DM and gas). We initialize the simulation at $z=100$ deep in the linear regime, adopting for this purpose a concordance $\Lambda$CDM cosmology with the following parameters: matter density $\Omega_{m}=1-\Omega_{\Lambda}=0.3$, baryon density $\Omega_b=0.04$, Hubble parameter $h=H_{0}/\left(100~\rm{km}~\rm{s}^{-1}~\rm{Mpc}^{-1}\right)=0.7$, spectral index $n_{\rm{s}}=1.0$, and a top-hat fluctuation power $\sigma_{8}=0.9$ \citep[e.g.,][]{spergel03}. Initial density and velocity perturbations are imprinted according to a Gaussian random field and grow in proportion to the scale factor until the onset of nonlinearity. At this point the detailed chemical evolution of the gas becomes crucial, and we apply the same chemical network as in \citet{jgb07} to track the abundances of H, H$^{+}$, H$^{-}$, H$_{2}$, H$_{2}^{+}$, He, He$^{+}$, He$^{++}$, and e$^{-}$, as well as the five deuterium species D, D$^{+}$, D$^{-}$, HD, and HD$^{-}$. All relevant cooling mechanisms in the temperature range $10$--$10^{8}~\rm{K}$ are implemented, including H and He resonance processes, bremsstrahlung, inverse Compton (IC) scattering, and molecular cooling for H$_{2}$ and HD. Metal cooling does not become important for the entire lifetime of the SN remnant, yet we postpone a more detailed discussion of this issue to \S5. We do not take into account the emission of radiation by the post-shock gas, which acts to create a thin layer of fully ionized material ahead of the shock and suppresses molecule formation \citep[e.g.,][]{sm79,sk87,ks92}, since (1) the SN remnant expands into an H~{\sc ii} region, and (2) we find that molecule formation in the post-shock gas becomes important only at late times, when it has cooled to below $10^{4}~\rm{K}$ (see \S3.4).

With these ingredients, the first star forms in a halo of $M_{\rm{vir}}\simeq 5\times 10^{5}~M_{\odot}$ and $r_{\rm{vir}}\simeq 100~\rm{pc}$ at $z\simeq 20$ in the canonical fashion \citep[e.g.,][]{bcl99,bcl02,abn02}. We determine its location by identifying the first particle that reaches a density of $n_{\rm{H}}=10^{4}~\rm{cm}^{-3}$. At this point the gas ``loiters'' around a temperature of $200~\rm{K}$ and typically attains a Jeans mass of a few $10^{3}~M_{\odot}$ before further collapsing \citep[e.g.,][]{bcl02,glover05}. For simplicity, we assume that such a clump forms a single star, and we find that its location is reasonably well resolved by the minimum resolution mass, $M_{\rm{res}}\simeq 500~M_{\odot}$. In Figure 3, we show the hydrogen number density in the x-y and y-z plane, centered on the formation site of the first star. Evidently, the host halo is part of a larger overdensity that will collapse in the near future and lead to multiple merger events. This behavior is characteristic of bottom-up structure formation, and our simulation therefore reflects a cosmological environment typical for these redshifts.

\begin{figure}
\begin{center}
\includegraphics[width=8.0cm,height=8.0cm]{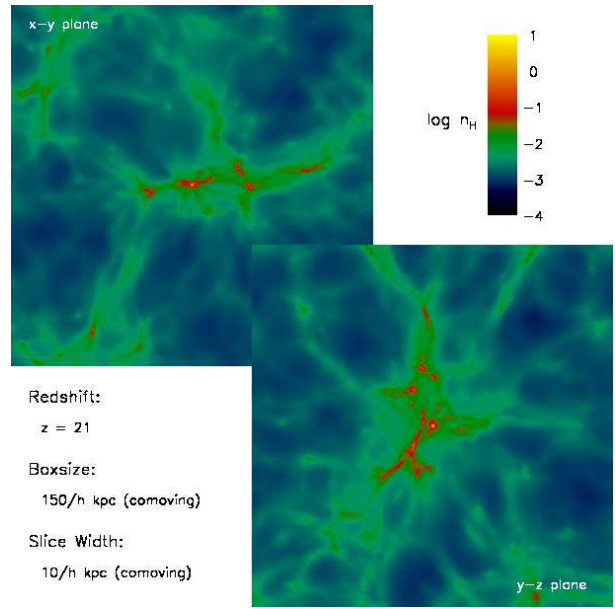}
\caption{Hydrogen number density averaged along the line of sight in a slice of $10/h~\rm{kpc}$ (comoving) around the first star, forming in a halo of total mass $M_{\rm{vir}}\simeq 5\times 10^{5}~M_{\odot}$ at $z\simeq 20$. Evidently, the host halo is part of a larger conglomeration of less massive minihalos and is subject to the typical bottom-up evolution of structure formation.}
\end{center}
\end{figure}

\subsubsection{H~{\sc ii} Region}
The treatment of the H~{\sc ii} region around the star is crucial for the early- and late-time behavior of the SN remnant. The photoevaporation of the host minihalo greatly reduces the central density and extends the energy-conserving ST phase, whereas after an intermediate stage the enhanced pressure in the H~{\sc ii} region leads to an earlier transition to the final, momentum-conserving phase. In addition, the shock fulfills the stalling criterion (i.e. $\dot{r}_{\rm{sh}}=c_{s}$, where $c_{s}$ is the sound speed of the photoheated IGM) much earlier in the H~{\sc ii} region compared to previously unheated gas. We have found that neglecting the presence of the H~{\sc ii} region around the star, which extends well into the IGM, leads to a final shock radius a factor of $2$ larger, demonstrating its importance for the long-term evolution of the SN remnant.

To determine the size and structure of the H~{\sc ii} region, we proceed analogously to \citet{jgb07}. In detail, we initially photoheat and photoionize a spherically symmetric region surrounding the star up to a maximum distance of $200~\rm{pc}$, where we find a neighboring minihalo. We determine the necessary heating and ionization rates by using the properties of a $200~M_{\odot}$ Pop~III star found by \citet{bkl01} and \citet{schaerer02}. After about $2~\rm{Myr}$, when the star has reached the end of its lifetime, the hydrodynamic shock has propagated to $r_{\rm{vir}}/2$ and photoevaporated the central regions of the host halo. Figure 8{\it a} shows the resulting density, temperature, pressure, and velocity profiles, which display the characteristics of the analytic \citet{shu02} solution (i.e. pressure equilibrium throughout the interior, while the density [temperature] become almost constant at small [large] radii). In our case, the average interior density drops to $n_{\rm{H}}\simeq 0.5~\rm{cm}^{-3}$, while the central temperatures rise to roughly $4\times 10^{4}~\rm{K}$ (see Fig. 8{\it a}). The radial profiles agree relatively well with the self-consistent radiation-hydrodynamics simulations performed in \citet{awb07} and \citet{yoshida07}, with slight differences most likely due to the harder spectrum and shorter lifetime of a $200~M_{\odot}$ star compared to a $100~M_{\odot}$ star \citep[see also][]{jb07}.

Since ionizing radiation escapes the minihalo after only a few thousand years \citep[e.g.,][]{abs06}, we determine the final size and structure of the H~{\sc ii} region by performing the ray-tracing algorithm introduced in \citet{jgb07}. This routine finds the Str\"{o}mgren radius along $10^{5}$ rays around the star, with each ray consisting of $200$ radial bins. For this purpose we assume that helium has the same ionization properties as hydrogen, and we do not treat a separate He~{\sc iii} front. Once an individual cell fulfills the Str\"{o}mgren criterion, we ionize its content and set its temperature and molecule abundances to the values at the outer edge of the photoheated region. This method is appropriate for the propagation of ionization fronts in the general IGM, but it does not correctly treat the photoevaporation of neighboring minihalos, when the ionization front becomes D type \citep[e.g.,][]{as07}. This issue is particularly important with respect to feedback on neighboring minihalos and is further discussed in \S4.

Figure 4 shows the gas temperature after the main-sequence lifetime of the star, and thus indirectly the size and structure of the H~{\sc ii} region. In agreement with \citet{abs06}, we find that the H~{\sc ii} region can be as large as $5~\rm{kpc}$, but efficient shielding by neighboring minihalos may limit its size to a few $100~\rm{pc}$ in some directions. The H~{\sc ii} region initially cools via IC scattering, while the pressure gradient at the boundary of the H~{\sc ii} region leads to a gradual adiabatic expansion on timescales comparable to the Hubble time at $z\simeq 20$ (see Figs. 4 and 9; also see Johnson \& Bromm 2007). Even though molecule fractions rise to typically $x_{\rm{H}_{2}}\sim 10^{-3}$ and $x_{\rm{HD}}\sim 10^{-7}$, molecular cooling remains inefficient due to the low densities of the photoheated gas \citep[e.g.,][]{jb07}. We do not treat the evolution of a separate Lyman-Werner (LW) front, as molecules inside the H~{\sc ii} region are destroyed primarily by collisional dissociation and charge transfer, while they are quickly reformed in more massive, neighboring minihalos once the central source turns off \citep[see][]{jgb07,yoshida07}.

\begin{figure}
\begin{center}
\includegraphics[width=8.0cm,height=8.0cm]{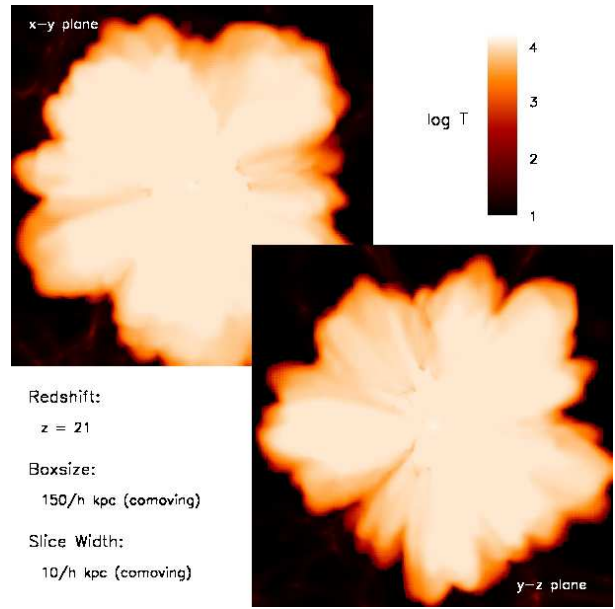}
\caption{Temperature averaged along the line of sight in a slice of $10/h~\rm{kpc}$ (comoving) around the star after its main-sequence lifetime of $2~\rm{Myr}$. Ionizing radiation has penetrated nearby minihalos and extends up to $5~\rm{kpc}$ around the source, heating the IGM to roughly $2\times 10^{4}~\rm{K}$, while some high-density regions have effectively shielded themselves.}
\end{center}
\end{figure}

\subsubsection{Sink Particles}
In the course of the simulation, neighboring minihalos might reach high enough densities to form stars, and the subsequent photoheating could significantly influence the propagation of the SN remnant. In the present work, we aim to investigate a single unperturbed SN explosion, and thus we rule out further star formation by employing the sink particle algorithm used in \citet{jgb07}. This routine forms sink particles once the hydrogen number density exceeds $10^{4}~\rm{cm}^{-3}$ and further accretes particles inside the Bondi radius \citep{bondi52}. This procedure prevents disturbances arising from the expansion of additional ionization fronts and allows us to concentrate on the feedback caused by the SN explosion.

With these preparations in place, we reinitialize the simulation at the end of the FE phase according to \S2.2. In the following we discuss the evolution of the SN remnant until it effectively dissolves into the IGM.

\section{Expansion and Cooling Properties}
Throughout its lifetime, the remnant goes through four evolutionary stages, each of which is characterized by a different physical mechanism becoming dominant. Based on this chronological sequence, we discuss the expansion and cooling properties of the SN remnant with respect to the simulation results, and we summarize the remnant's behavior with a simple analytic model.

\subsection{Phase I: Free Expansion}
At very early times, the SN remnant enters the FE phase and propagates nearly unhindered into the surrounding medium. It expands with a constant velocity of $v_{\rm{ej}}^{2}=2E_{\rm{sn}}/M_{\rm{ej}}$, where for our case $M_{\rm{ej}}=M_{*}$. The duration of the FE phase is given by $t_{\rm{fe}}=r_{\rm{fe}}/v_{\rm{ej}}$, or
\begin{equation}
t_{\rm{fe}}=r_{\rm{fe}}\sqrt{\frac{M_{\rm{ej}}}{2E_{\rm{sn}}}}\mbox{\ ,}
\end{equation}
where $r_{\rm{fe}}$ is the radius at which the swept-up mass equals the mass of the original ejecta; i.e,
\begin{equation}
r_{\rm{fe}}=\left(\frac{3XM_{\rm{ej}}}{4\pi m_{\rm{H}}n_{\rm{H}}}\right)^{1/3}\mbox{\ ,}
\end{equation}
where $X=0.76$ is the primordial mass fraction of hydrogen \citep[e.g.,][]{ky05}.

After $t_{\rm{fe}}$, the inertia of the swept-up mass becomes important, and the shock undergoes a transition to the ST phase. Since we do not explicitly model the FE phase, we use the above analytic arguments to obtain $r_{\rm{fe}}\lesssim 20~\rm{pc}$ and $t_{\rm{fe}}\lesssim 10^{4}~\rm{yr}$ (see Fig. 7).

\subsection{Phase II: Sedov-Taylor Blast Wave}
According to \S2.2, the simulation begins when the shock undergoes a transition to the ST phase. Since we have already discussed the properties of the ST solution in \S2.3, we can apply these results to understand the early behavior of the SN remnant. With the knowledge that the previous photoheating has created an average density profile of $n_{\rm{H}}\simeq0.5~\rm{cm}^{-3}$ in the vicinity of the progenitor (Fig. 8{\it a}), we find with the help of equation (5) that the shock approaches $r_{\rm{vir}}/2$ after about $10^{5}~\rm{yr}$. At this point it catches up with the previously established photoheating shock, where the outlying density profile becomes isothermal. The simulation results (Fig. 8{\it b}) agree with this analytic prediction, yet the profiles do not quite resemble the ST solution, partly because not enough time has passed, but also because cooling in the dense shell becomes important, violating energy conservation. Consequently, the ST phase ends and the remnant undergoes a second transition.

To understand the origin of this transition, we elaborate on the cooling mechanisms responsible for radiating away the thermal energy of the remnant. At the high temperatures behind the shock, these are H and He resonance processes (RPs), bremsstrahlung (BS), and IC scattering, with the former consisting of collisional ionization, excitation, and recombination cooling. Figure 5 shows the relevant cooling rates, normalized to a hydrogen number density of unity, where the abundances of the various species follow from assuming collisional ionization equilibrium. Evidently, IC scattering and BS are important at high temperatures, whereas RPs become significant below $10^{6}~\rm{K}$. However, BS and RPs are proportional to the density squared, while IC scattering exhibits only a linear dependence. Thus, RPs will be the most important coolant in the dense shell, while IC scattering is dominant in the hot interior.

\begin{figure}
\begin{center}
\includegraphics[width=8.0cm,height=5.0cm]{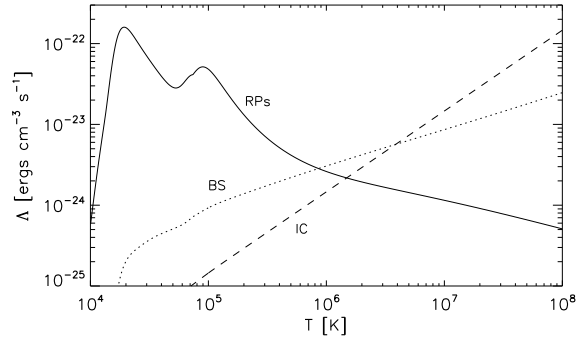}
\caption{Cooling rates as a function of temperature at $z\simeq 20$ for H and He RPs (solid line), BS (dotted line), and IC scattering (dashed line) in collisional ionization equilibrium, with $n_{\rm{H}}$ set to unity. At low temperatures RPs dominate, while above $10^{6}~\rm{K}$ IC scattering and BS become important. However, RP's and BS are proportional to the density squared, while IC scattering exhibits only a linear dependence.}
\end{center}
\end{figure}

At $z=20$, the cooling time for IC scattering is approximately $10~\rm{Myr}$, independent of temperature and density, while cooling times for RPs in the dense shell are generally much lower. To determine when radiative losses affect the energetics of the SN remnant, we must equate the shell cooling time with the expansion time. Figure 6 shows this analytic relation for the initial conditions of the simulation, i.e. $E_{\rm{sn}}=10^{52}~\rm{ergs}$ and $n_{\rm{H}}=0.5~\rm{cm}^{-3}$. Confirming the simulation results, this prediction yields that RPs efficiently cool the dense shell to $10^{4}~\rm{K}$ after about $10^{5}~\rm{yr}$. At this point the shocked gas separates into a hot, interior bubble with temperatures above $10^{6}~\rm{K}$ and a dense shell at $10^{4}~\rm{K}$ bounded by a high-pressure gradient. This multi-phase structure is clearly visible in Figure 8{\it b}, which remains intact for $\la 10~\rm{Myr}$, when IC scattering becomes important and cools the last remnants of the interior bubble to the temperature of the dense shell. With energy conservation no longer valid, the ST phase ends before the shock has relaxed to the asymptotic solution, yet we nevertheless find that $r_{\rm{sh}}\propto t_{\rm{sh}}^{2/5}$ fits the temporal scaling of the mass-weighted mean shock radius relatively well (see Fig. 7).

\begin{figure}
\begin{center}
\includegraphics[width=8.0cm,height=5.0cm]{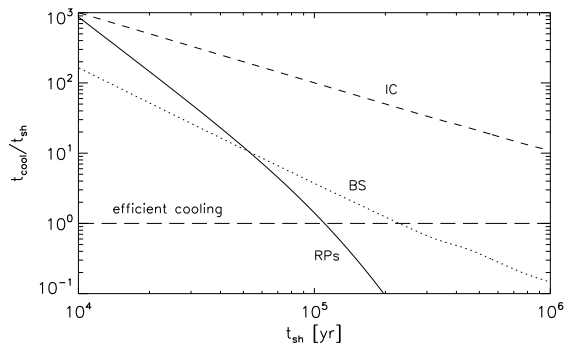}
\caption{Cooling time over expansion time in the dense shell for RPs (solid line), BS (dotted line), and IC (dashed line). The long-dashed line indicates when cooling becomes efficient, implying that RPs are responsible for ending the ST phase after $10^{5}~\rm{yr}$. At $z=20$, the cooling time for IC scattering is roughly $10~\rm{Myr}$, independent of temperature and density.}
\end{center}
\end{figure}

\subsection{Phase III: Pressure-Driven Snowplow}
After the ST phase ends, the pressurized, interior bubble drives a dense shell, and one speaks of a pressure-driven snowplow (PDS). To analytically describe the further evolution of the SN remnant, we assume that the entire swept-up mass $M_{\rm{sw}}$ is confined to an infinitely thin shell. In light of the steep density profile toward the interior, this assumption is justified, leading to an equation of motion of the form:
\begin{equation}
\frac{d\left(M_{\rm{sw}}v_{\rm{sh}}\right)}{dt}=4\pi r_{\rm{sh}}^{2}P_{b}\mbox{\ ,}
\end{equation}
where $P_{b}$ is the pressure of the hot, interior bubble, and the external pressure has been neglected \citep[e.g.,][]{om88}. Since $M_{\rm{sw}}\propto r_{\rm{sh}}$ in an $r^{-2}$ density profile and $P_{b}\propto r_{\rm{sh}}^{-5}$ in the adiabatically expanding interior, one can solve the above equation with a power law of the form $r_{\rm{sh}}\propto t_{\rm{sh}}^{2/5}$. Interestingly, this procedure yields the same temporal scaling as the ST solution, and we therefore do not expect a change in slopes after $10^{5}~\rm{yr}$. This is a direct consequence of the transition to the PDS phase once the shock approaches the isothermal density profile in the outskirts of the halo. Figure 7 confirms this result, as the simulation shows only a slight deviation from the analytically derived $t_{\rm{sh}}^{2/5}$ slope.

Using the above model, we can further determine when the PDS phase ends. The pressure directly behind the shock after $10^{5}~\rm{yr}$ can be estimated with equation (3) as $P_{b}/k_{\rm{B}}\simeq 3\times 10^{6}~\rm{K}~\rm{cm}^{-3}$, in agreement with the simulation (see Fig. 8{\it b}). With $P_{b}\propto r_{\rm{sh}}^{-5}$ and $r_{\rm{sh}}\propto t_{\rm{sh}}^{2/5}$, we further find $P_{b}\propto t_{\rm{sh}}^{-2}$, implying that after roughly $1~\rm{Myr}$ the interior pressure has dropped to $P_{\rm{b}}/k_{\rm{B}}\simeq 3\times 10^{4}~\rm{K}~\rm{cm}^{-3}$. At this point pressure equilibrium between the hot interior and the dense shell has been established, and the shock is driven solely by its accumulated inertia. Figure 8{\it c} confirms this prediction, showing that the interior pressure has indeed dropped to that of the dense shell. In contrast, the temperature only drops proportional to $t_{\rm{sh}}$ and remains high.

\subsection{Phase IV: Momentum-Conserving Snowplow}
With the pressure gradient no longer dominant, the SN remnant is driven by the accumulated inertia of the dense shell and becomes a momentum-conserving snowplow (MCS). In analogy to the derivation performed in \S3.3, the position of the shock as a function of time can be obtained by solving equation (8) in the absence of a pressure term. Since the shock has not yet propagated beyond the surrounding $r^{-2}$ density profile, this yields an initial scaling of $r_{\rm{sh}}\propto t_{\rm{sh}}^{1/2}$.

At later times, the shock finally leaves the host halo and encounters neighboring minihalos in the y-z plane, but underdense voids perpendicular to the y-z plane (see Fig. 3). This increases its radial dispersion, while at the same time the Hubble expansion becomes important and serves to expand the medium on which the SN remnant propagates, thus raising its physical shock velocity. We can therefore only estimate the temporal scaling of the mass-weighted mean shock radius based on the simulation results, finding, interestingly, that $r_{\rm{sh}}$ maintains its $t_{\rm{sh}}^{1/2}$ scaling until it fulfills the stalling criterion (see Fig. 7). This occurs after about $200~\rm{Myr}$, close to the Hubble time at $z\simeq 20$, when the shock velocity approaches the local IGM sound speed and becomes indistinguishable from sound waves. Evidently, the increased slope at late times is a fundamental difference between SNe occurring at high redshifts and in the present-day universe, where $r_{\rm{sh}}\propto t_{\rm{sh}}^{1/4}$.

In the meantime, the hot interior has expanded adiabatically for roughly $10~\rm{Myr}$, when IC scattering becomes important, quickly cooling the interior to temperatures of the dense shell (see Fig. 8{\it d}). Even later, the high electron fraction persisting in the interior and the dense shell leads to efficient molecule formation, and typical abundances of $x_{\rm{H}_{2}}\sim 10^{-3}$ and $x_{\rm{HD}}\sim 10^{-7}$ are reformed. By the end of the simulation, both phases have adiabatically cooled to $T\sim 10^{3}~\rm{K}$, while densities in the dense shell have dropped to $n_{\rm{H}}\sim 10^{-2.5}~\rm{cm}^{-3}$ (see Figs. 9 and 14). The interior is even more underdense, and we find that the SN explosion has completely disrupted the host halo, preventing further star formation inside of it for at least a Hubble time at $z\simeq 20$. Due to the continuous adiabatic expansion of the post-shock gas, molecular cooling remains inefficient for the entire lifetime of the SN remnant.

\subsection{Summary of Expansion Properties}
Summarizing the expansion properties of the SN remnant, Figure 7 shows the mass-weighted mean shock radius as a function of time, together with the analytically derived power laws. The good agreement indicates that one can quantify the evolution of the SN remnant by means of simple physical arguments and obtain relatively accurate results.

We find a final mass-weighted mean shock radius of $2.5~\rm{kpc}$, which is roughly a factor of $2$ smaller than the H~{\sc ii} region. Figure 7 also shows the radial dispersion of the shock, indicating that the surrounding medium becomes highly anisotropic once the SN remnant leaves the host halo. After about $5~\rm{Myr}$, the shock encounters the first neighboring minihalos in the y-z plane and begins to stall, while it continues to propagate into the voids perpendicular to the y-z plane. The vast majority of the shocked material resides in the general IGM, as the mass-weighted mean closely traces the maximum shock radius.

To conclude this section, Figure 10 shows the swept-up gas mass as a function of time for the simulation and the analytic model, where the slopes have been determined by the relation $M_{\rm{sw}}\propto \rho r_{\rm{sh}}^{3}$, except at very late times when the shock structure becomes too complex for a simple analysis. We find a final swept-up mass of $2.5\times 10^{5}~M_{\odot}$, which is roughly a factor of $2$ smaller than the mass enclosed inside a sphere of radius $2.5~\rm{kpc}$, assuming the background density at $z\simeq 12$. This further demonstrates the overall asymmetry of the shock (see Fig. 14).

\begin{figure}
\begin{center}
\includegraphics[width=8.0cm,height=8.0cm]{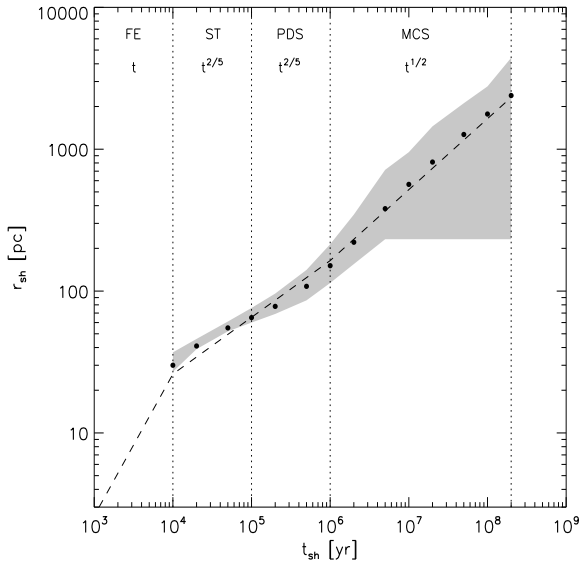}
\caption{Evolution of the SN remnant in its entirety, from $z\simeq 20$, when the SN explodes, to $z\simeq 12$, when the shock finally stalls. Its total lifetime is about $200~\rm{Myr}$, or a Hubble time at $z\simeq 20$. The black dots indicate the mass-weighted mean shock radius according to the simulation, while the dashed line shows the analytic solution. For both we find a final mass-weighted mean shock radius of $2.5~\rm{kpc}$. The shaded region shows the radial dispersion of the shock, indicating that it increases significantly once the SN remnant leaves the host halo and encounters the first neighboring minihalos in the y-z plane. The bulk of the SN remnant propagates into the IGM, since the mass-weighted mean closely traces the maximum shock radius.}
\end{center}
\end{figure}

\begin{figure*}
\begin{center}
\resizebox{16cm}{16cm}
{\includegraphics{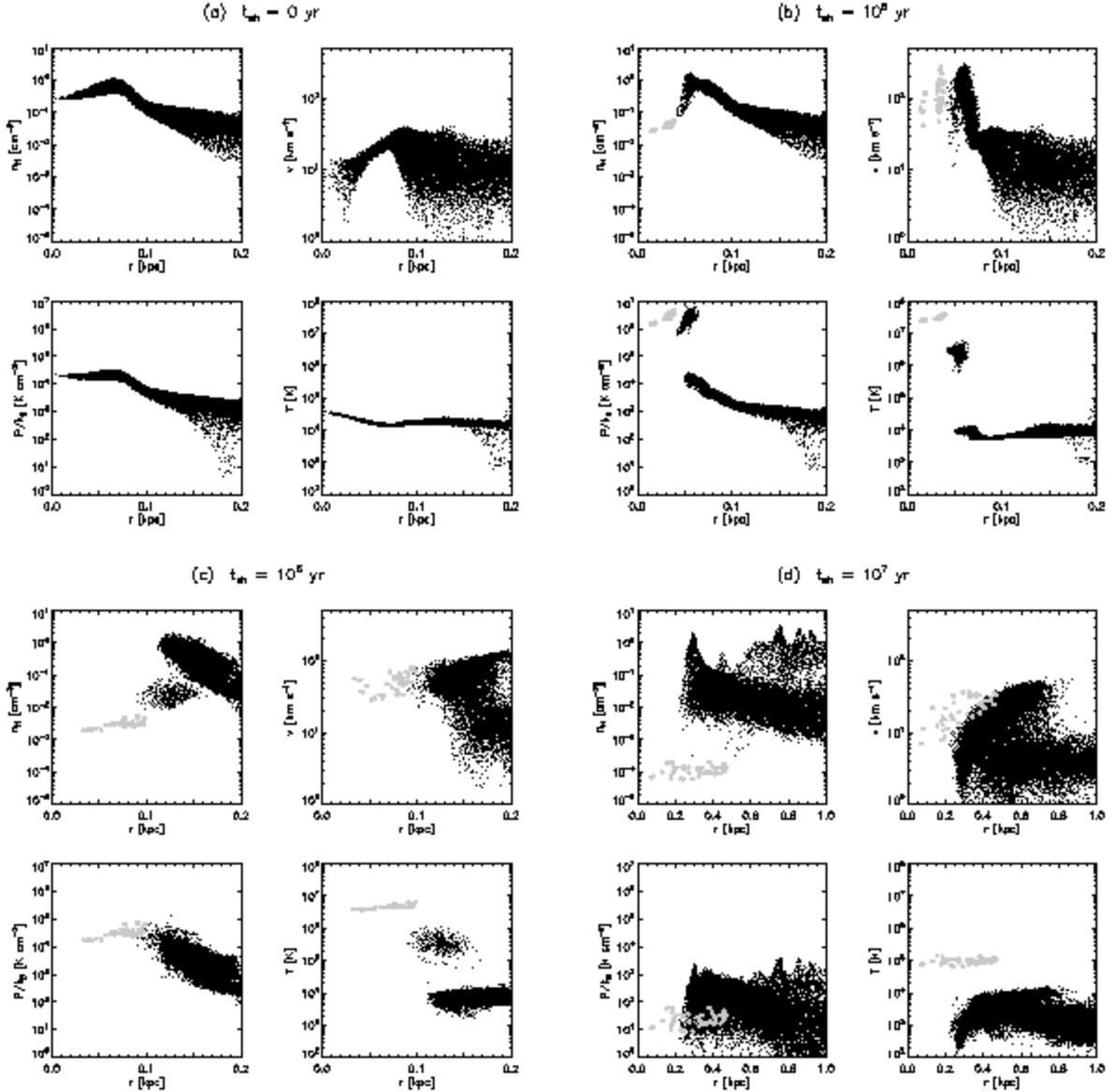}}
\caption{Density, velocity, pressure, and temperature around the SN progenitor star immediately before its death, and $10^{5}$, $10^{6}$, and $10^{7}~\rm{yr}$ after the SN explosion. Black dots represent normal SPH particles, while gray dots represent the initial stellar ejecta. {\it (a)}: The hydrodynamic shock of the H~{\sc ii} region has approached $r_{\rm{vir}}/2$ at $\simeq 50~\rm{pc}$ after establishing pressure equilibrium in the interior, while the density and velocity profiles assume the characteristic \citet{shu02} solution of a champagne flow. The temperature of the H~{\sc ii} region is typically $2\times 10^{4}~\rm{K}$. {\it (b)}: The shock created by the SN explosion propagates into the surrounding medium according to the ST solution, with the characteristic formation of a dense shell and a hot, interior bubble. After $10^{5}~\rm{yr}$, the shock approaches the previous photoheating shock with expansion velocities well in excess of $100~\rm{km}~\rm{s}^{-1}$. At the same time, cooling in the dense shell by RPs becomes efficient and the ST phase ends. The resulting sharp temperature drop between the interior bubble at $T>10^{6}~\rm{K}$ and the dense shell at $T\simeq 10^{4}~\rm{K}$ is clearly visible. {\it (c)}: The SN remnant has undergone a transition to the PDS phase and is driven by the high pressure in the interior and the momentum of the dense shell. This phase ends after about $1~\rm{Myr}$, when pressure equilibrium has been established. The pressure drop in comparison with {\it (b)} is clearly visible, while temperatures in the interior remain high. {\it (d)}: After $10~\rm{Myr}$, the radial dispersion of the shock has increased dramatically, ranging from $\ga 200~\rm{pc}$ where the shock encounters the first neighboring minihalos, to $\la 1~\rm{kpc}$ perpendicular to the overdensities in the y-z plane.}
\end{center}
\end{figure*}

\begin{figure*}
\begin{center}
\resizebox{16cm}{16cm}
{\includegraphics{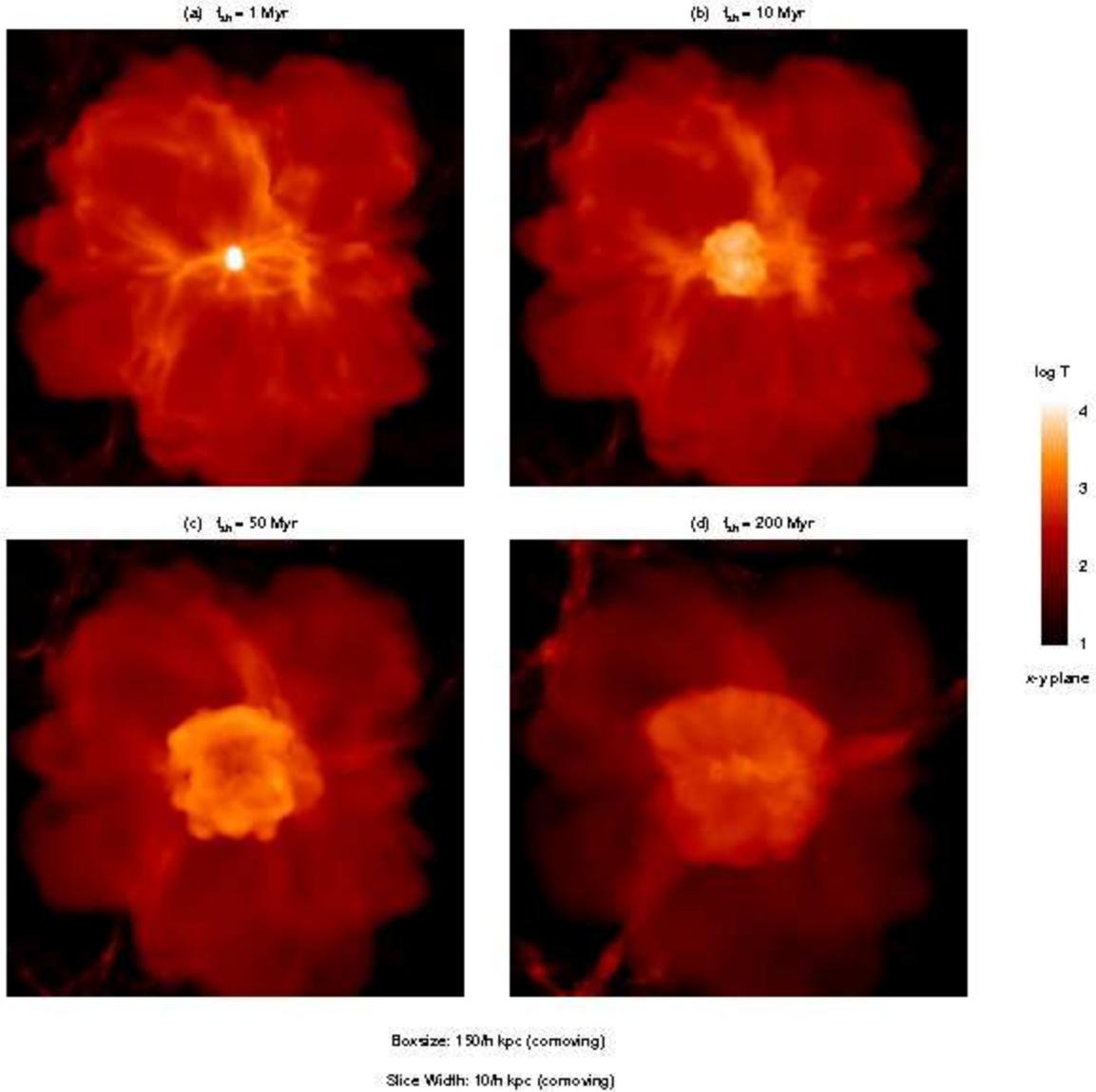}}
\caption{Temperature averaged along the line of sight in a slice of $10/h~\rm{kpc}$ (comoving) around the x-y plane after $1$,  $10$, $50$, and $200~\rm{Myr}$. In all four panels, the H~{\sc ii} region and SN shock are clearly distinguishable, with the former occupying almost the entire simulation box, while the latter is confined to the central regions. {\it (a)}: The SN remnant has just left the host halo, but temperatures in the interior are still well above $10^{4}~\rm{K}$. {\it (b)}: After $10~\rm{Myr}$, the asymmetry of the SN shock becomes visible, while most of the interior has cooled to well below $10^{4}~\rm{K}$. {\it (c)}: The further evolution of the shocked gas is governed by adiabatic expansion, and its morphology develops a ``finger-like'' structure (see also Fig. 13). {\it (d)}: After $200~\rm{Myr}$, the shock velocity approaches the local sound speed and the SN remnant stalls. By this time the post-shock regions have cooled to roughly $10^{3}~\rm{K}$.}
\end{center}
\end{figure*}

\begin{figure}
\begin{center}
\includegraphics[width=8.0cm,height=5.0cm]{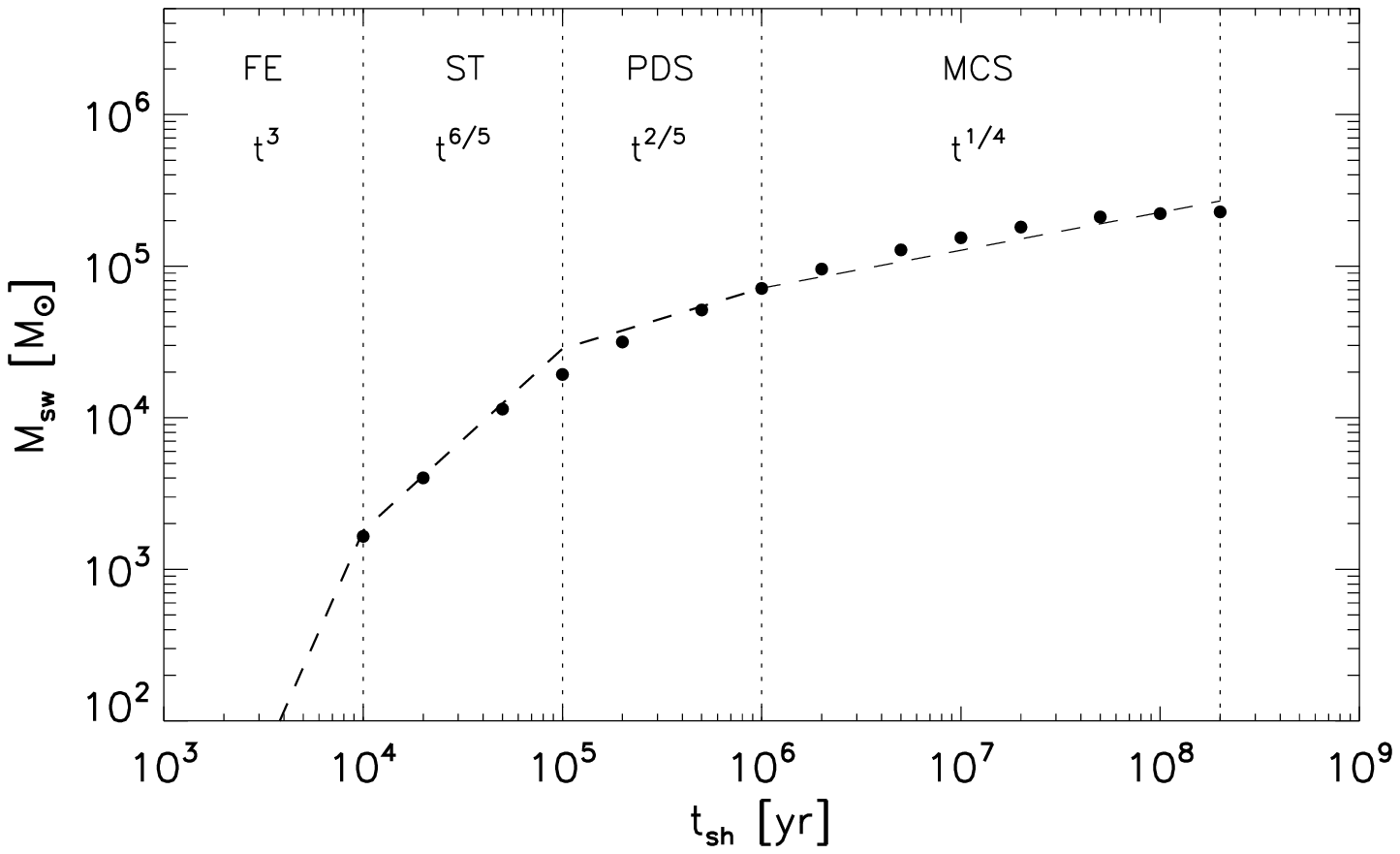}
\caption{Swept-up gas mass as a function of time for the simulation (black dots) and the analytic model (dashed lines), where the latter is related to the shock radius by $M_{\rm{sw}}\propto\rho r_{\rm{sh}}^{3}$, except at very late times, when the shock structure becomes too complex for a simple analysis. For both we find a final swept-up mass of $2.5\times 10^{5}~M_{\odot}$.}
\end{center}
\end{figure}

\section{Feedback on Neighboring Halos}
So far we have concentrated on the propagation of the SN remnant into the IGM and treated neighboring minihalos as disturbances to the radial density profile. In the following, we explicitly investigate the mechanical feedback of the SN on nearby minihalos and discuss possible consequences for star formation. For this purpose we have carried out two additional simulations, one without feedback and another with radiative feedback, which enable us to disentangle the effects of photoheating and the SN shock.

Combining the so-obtained results, Figure 11 shows the distances of all star-forming halos from the initial SN progenitor as a function of expansion time, when they have reached the threshold density $n_{\rm{H}}=10^{4}~\rm{cm}^{-3}$. Following our argumentation in \S2.4, the halo irreversibly collapses at this point and forms a star. The shades of the symbols in Figure 11 indicate their affiliation to the different simulations (i.e. black, dark gray, and light gray symbols represent the no-feedback, photoheating-only, and main simulation runs), while the shapes of the symbols represent individual halos. For orientation, we show the mass-weighted mean shock radius at late times according to Figure 7.

\subsection{Delay by Photoheating}
In the photoheating-only case, we find that the collapse of all three star-forming halos is delayed (see Fig. 11). While we do not properly resolve D-type ionization fronts that may develop within shielded minihalos, we find that ionizing radiation can penetrate deep into their cores and suppress cooling and the accretion of gas. The distance, mass, and maximum density of the halo at the onset of photoheating are all crucial for the extent of the delay. The most massive halo in our simulation, halo a, experiences a delay of only $25~\rm{Myr}$, while halos b and c are less dense at the time of photoheating and experience delays in excess of $80~\rm{Myr}$. This is roughly consistent with the results of \citet{as07} who argue that radiative feedback on neighboring minihalos strongly depends on their evolutionary stage \citep[see also][]{mbh06,su06}. For more quantitative results, one must perform self-consistent radiation-hydrodynamics simulations that include the effects of photodissociating radiation \citep[e.g.,][]{yoshida07}.

Due to the inaccuracies mentioned above, we cannot draw robust conclusions on the subsequent mode of primordial star formation, i.e., Pop~III or Pop~II.5, but we anticipate this to be a function of the state of the collapse and the ionizing flux \citep[see also][]{mbh03,gb06,jb06,jappsen07}.

\subsection{Shock-driven Collapse}
In contrast to feedback by ionizing radiation, we find that in our case the shock of the SN remnant acts to enhance halo collapse and promote star formation. This behavior is evident from Figure 11, as halos a and b collapse about $15~\rm{Myr}$ earlier with respect to the photoheating-only run. In both cases, the shock compresses the dense cores of each of the halos, which serves to enhance cooling and accretion. Although our resolution is too crude for quantitative conclusions, we find that the mechanical impact of the SN remnant mitigates the delay caused by photoheating and leads to a slightly increased star formation rate. This situation might be different for star-forming halos much closer to the SN explosion, where ram pressure stripping could be sufficient to dispel a large fraction of the gas. However, in the present simulation we find that most halos massive enough to form stars are at sufficient distances from the SN progenitor.

We note that we neglect the possible effects of photodissociating radiation emanating from the relic H~{\sc ii} region, as it has been shown that recombination radiation can only produce a relatively weak photodissociating flux, which quickly dies away as the gas in the relic H~{\sc ii} region recombines on timescales of the order of $1~\rm{Myr}$ (e.g., Johnson \& Bromm 2007; Yoshida et al. 2007; see also Glover 2007). In addition, we are able to neglect the radiation generated by the supernova-shocked gas as a source of photodissociating radiation concerning neighboring minihalos, since the shock is only moving at velocities $\la 20~\rm{km}~\rm{s}^{-1}$ when it arrives at the star-forming minihalos mentioned above. This implies a photodissociating flux of $J_{\rm{LW}}\la 10^{-5}$, where $J_{\rm{LW}}$ is in units of $10^{-21}~\rm{ergs}~\rm{s}^{-1}~\rm{cm}^{-2}~\rm{Hz}^{-1}~\rm{sr}^{-1}$ \citep{ss79}. Such a low level of flux will not have a substantial impact on the evolution of gas in collapsing minihalos, as it implies a timescale for photodissociation of H$_2$ of the order of $1~\rm{Gyr}$, much longer than the Hubble time at the redshifts we consider \citep[e.g.,][]{oh03,jb06}. However, a minihalo within $\sim 200~\rm{pc}$ of the SN explosion may be subject to a significant photodissociating flux, owing to the higher shock velocity at small distances.

\subsection{Mixing Efficiency}
An important question is whether a large fraction of metals can penetrate neighboring halos and efficiently mix with their cores, thus changing the mode of star formation from Pop~III/Pop~II.5 to Pop~II. Although a detailed quantitative analysis would require dedicated high-resolution simulations that are unavailable here, we can nevertheless estimate the mixing efficiency by applying the criterion for the operation of Kelvin-Helmholtz (KH) instabilities given by Murray et al. (1993; see also Wyithe \& Cen 2007; Cen \& Riquelme 2007):
\begin{equation}
\frac{g D r_{\rm{vir}}}{2\pi \dot{r}_{\rm{sh}}^2} \la 1\mbox{\ ,}
\end{equation}
where $g$ is the gravitational acceleration at the virial radius, and $D$ the density ratio of gas in the halo compared to the dense shell. For the nearest star-forming halo with about $5\times 10^{5}~M_{\odot}$ and $r_{\rm{vir}}\simeq 100~\rm{pc}$, we find $D\simeq 10$ and $\dot{r}_{\rm{sh}}\simeq 20~\rm{km}~\rm{s}^{-1}$. The left-hand side of equation (9) is thus of order $1/10$, suggesting that the gas at the outer edge of the halo will be disrupted and mix with the enriched material in the dense shell, while the core of the halo will remain pristine and stable. This is consistent with the results of the detailed simulations conducted by \citet{cr07}, and by the same arguments we find that halos b and c remain pristine as well.

Once again, we note that massive minihalos very close to the SN progenitor might experience a different behavior, as expansion velocities are of the order $100~\rm{km}~\rm{s}^{-1}$ when the SN remnant leaves the host halo (see Fig. 8). In such cases, the shock could disrupt the dense cores and trigger efficient mixing, leading to Pop~II star formation once the gas recollapses.

\subsection{Gravitational Fragmentation}
Various authors have suggested that cooling in the dense shell might lead to gravitational fragmentation and trigger secondary star formation \citep[e.g.,][]{mbh03,sfs04,machida05}. However, due to the previous photoheating, the density of the surrounding medium is sufficiently lowered such that molecule formation is initially inefficient. Even though we do not fully resolve the radiative shock front, we thus find that the continuous adiabatic expansion of the dense shell renders molecular cooling unimportant and does not trigger gravitational instabilities (see \S3.4 and Fig. 8). We note that the potential mixing of metals into the dense shell and the associated additional cooling might affect these results but postpone a more detailed discussion of this issue to future work.

Another scenario for secondary star formation is at the interface of colliding SN remnants. If such encounters take place in significantly overdense regions and occur relatively early in the evolution of individual SN remnants, densities might become high enough for gravitational fragmentation to occur.

\begin{figure}
\begin{center}
\includegraphics[width=8.0cm,height=7.5cm]{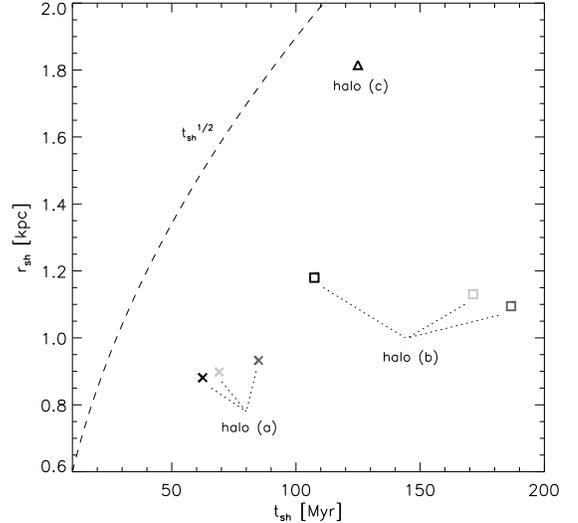}
\caption{Collapse times and distances from the SN progenitor for all star-forming minihalos affected by the SN shock. The shades of the symbols indicate their affiliation, i.e. black, dark gray, and light gray symbols represent the no-feedback, photoheating-only, and main simulation runs, respectively, while the shapes of the symbols denote the individual halos. For orientation, the dashed line shows the mass-weighted mean shock radius at late times according to Fig. 7. In our case, photoheating significantly delays star formation, while the SN shock acts to compress gas in neighboring minihalos and slightly accelerates their collapse.}
\end{center}
\end{figure}

\section{Chemical Enrichment}
We have previously argued that mixing of enriched material with gas in existing star-forming halos is generally inefficient (see \S4.3), indicating that the dispersal of metals can only occur via expulsion into the IGM. In \S3.5 we have shown that the bulk of the shock propagates into the voids surrounding the host halo, and we expect that chemical enrichment proceeds via the same channel \citep[e.g.,][]{pmg07}. This realization is crucial, as the detailed distribution of metals not only governs the transition to secondary star formation, but also the properties of the first galaxies, when the enriched material recollapses into larger potential wells.

We shed some light on these issues by determining the influence of metal cooling on the evolution of the SN remnant and by discussing the transport of metals into the IGM. Although our resolution is too crude for a detailed analysis, we can nevertheless discuss the mixing efficiency in a qualitative manner and draw some preliminary conclusions on the ultimate fate of the expelled metals.

\begin{figure}
\begin{center}
\includegraphics[width=8.0cm,height=5.0cm]{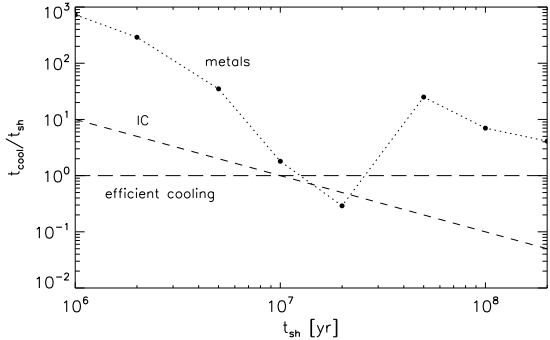}
\caption{Cooling time vs. expansion time in the interior. The dotted line represents cooling by metal lines for a total metallicity of $Z=Z_{\odot}$, but consisting solely of C, O, Fe, and Si (in equal parts), while the dashed line represents IC cooling. The long-dashed line indicates when cooling becomes efficient, implying that metal cooling becomes momentarily important after $10~\rm{Myr}$, when the interior temperatures have dropped to $10^{5}~\rm{K}$ and cooling rates peak. However, IC scattering becomes efficient at the same time and rapidly cools the enriched regions to below $10^{4}~\rm{K}$, where metal cooling rates drop by a few orders of magnitude. Consequently, the presence of metals does not effect the dynamical evolution of the SN remnant.}
\end{center}
\end{figure}

\begin{figure*}
\begin{center}
\resizebox{16cm}{13cm}
{\includegraphics{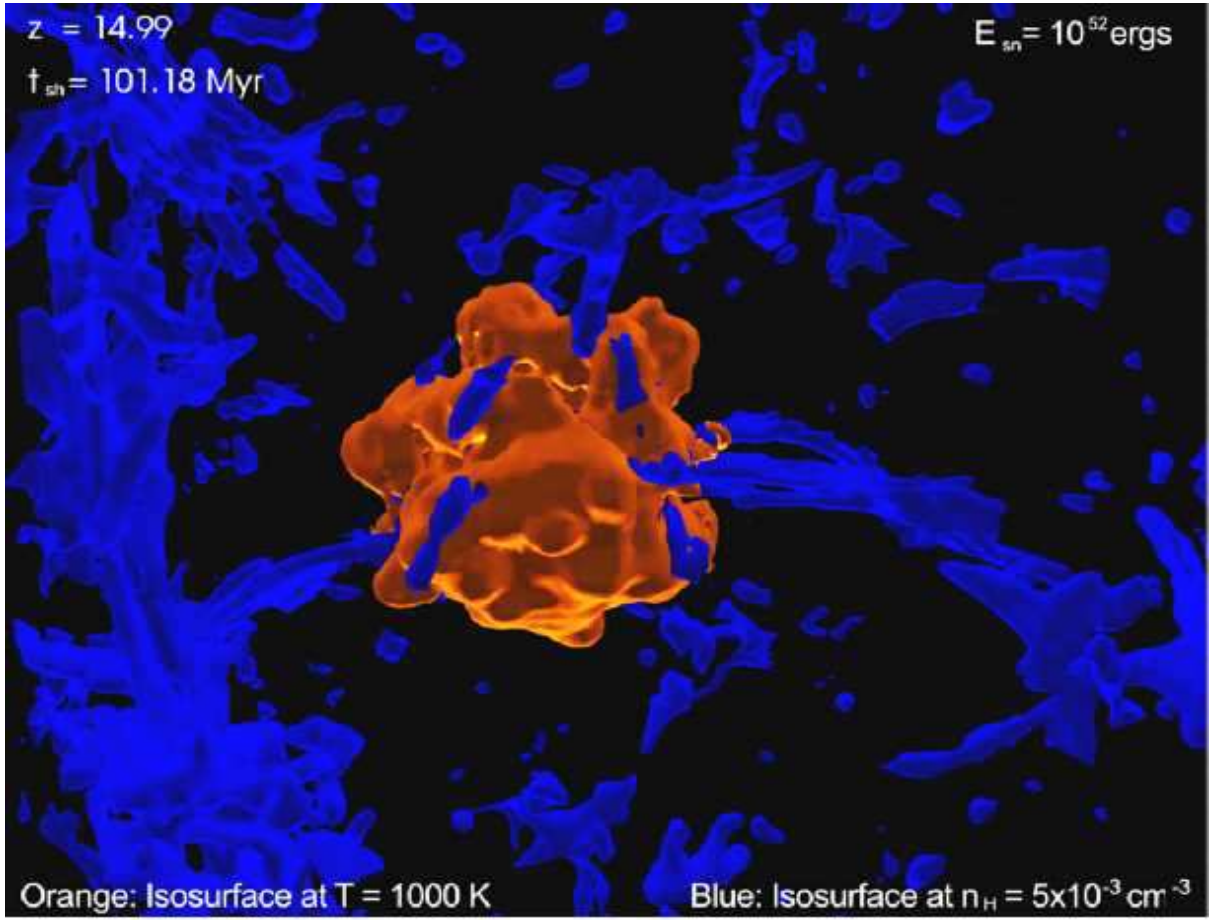}}
\caption{Three-dimensional view of the SN remnant at $z=15$, or $100~\rm{Myr}$ after the SN explosion, when it has reached a radius of $2~\rm{kpc}$. The finger-like morphology of the shock becomes visible as the SN remnant propagates at varying speeds in different directions, caused by an anisotropic density distribution.}
\end{center}
\end{figure*}

\begin{figure*}
\begin{center}
\resizebox{16cm}{16cm}
{\includegraphics{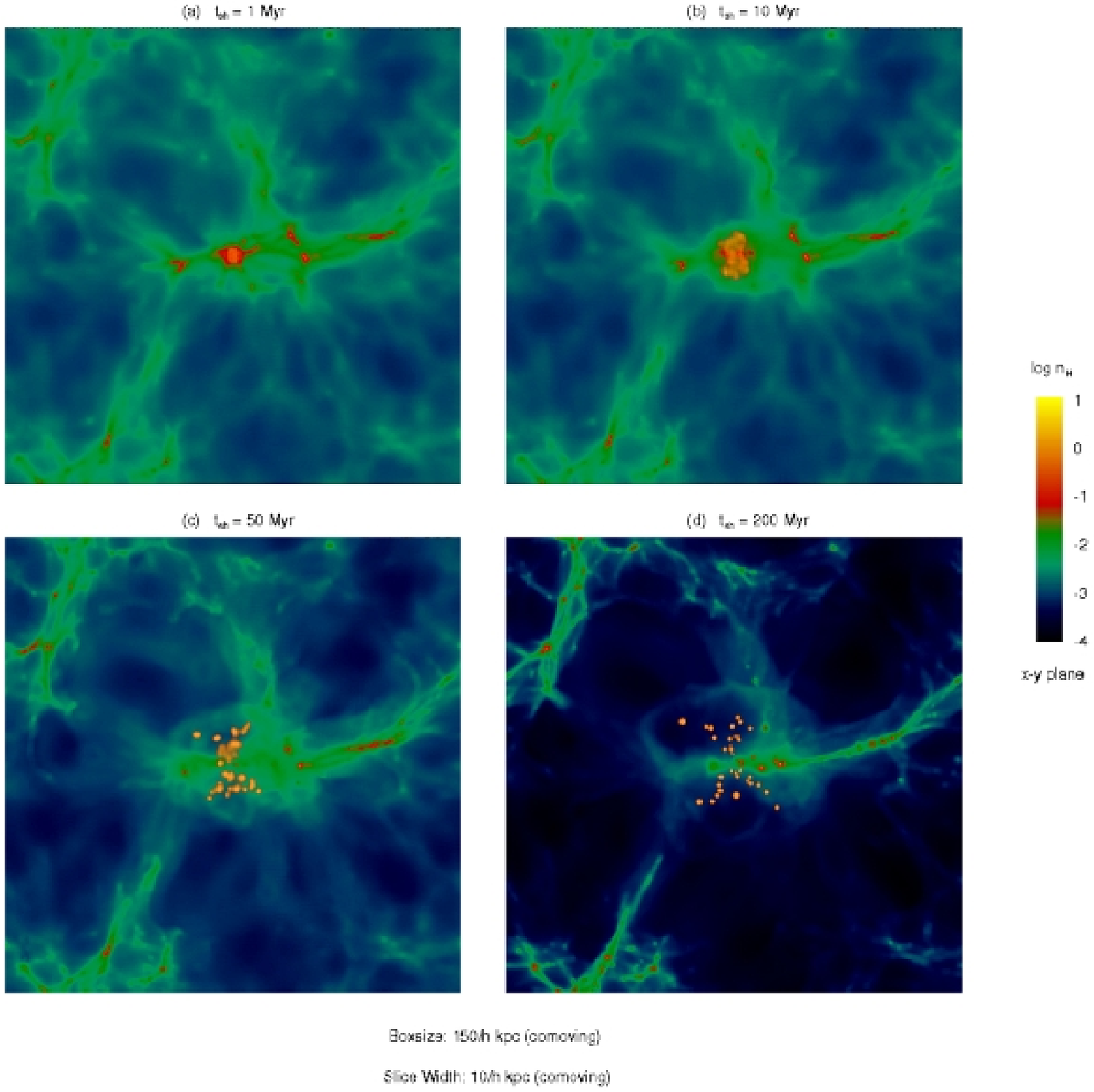}}
\caption{Hydrogen number density averaged along the line of sight in a slice of $10/h~\rm{kpc}$ (comoving) around the x-y plane, overlaid with the distribution of all metal particles (bright orange), after $1$,  $10$, $50$, and $200~\rm{Myr}$. The extent of each particle is set by the SPH smoothing length, while the apparent mixing within neighboring minihalos is a projection effect. {\it (a)}: After $1~\rm{Myr}$, the shock has left the host halo and enters the IGM, while the metals are still confined well within the virial radius. {\it (b)}: The contours of the shock have become visible as the SN remnant plows into the IGM $10~\rm{Myr}$ after the SN explosion. The interior bubble expands adiabatically into the cavities created by the shock and begins to lose its spherical symmetry. {\it (c)}: After $50~\rm{Myr}$, the dense shell becomes distinguishable from the metal-enriched interior, which has become substantially asymmetric and expands into the voids around the y-z plane in the shape of an hourglass. {\it (d)}: When the SN remnant finally stalls after $200~\rm{Myr}$, the dense shell and the metal-enriched, interior bubble have reached their maximum extent, with the former providing a natural confinement around the interior.}
\end{center}
\end{figure*}

\subsection{Heat Conduction and Metal Cooling}
During the first $10~\rm{Myr}$, the initial stellar ejecta expand adiabatically and apparently do not mix with the surrounding material (see Fig. 8). In reality, however, the high electron mean free path behind the shock leads to heat conduction and gas from the dense shell evaporates into the hot, interior bubble \citep[e.g.,][]{gull73}. This effect is not included in the simulation, as it would only marginally affect the dynamics of the SN remnant, but the onset of Rayleigh-Taylor (RT) and KH instabilities is predicted to efficiently mix the metals with primordial material evaporated from the dense shell, reducing the metallicity of the interior by a factor of a few to at most one order of magnitude \citep[e.g.,][]{mfr01}. In the following, we assume that the overall metallicity has dropped by a factor of $5$, to $Z=Z_{\odot}$, in the course of the first million years.

To determine the importance of metal cooling, we proceed analogously to \S3.2 and compare the expansion and cooling timescales of the interior bubble. For this purpose we adopt the cooling rates provided in \citet{maio07} for gas in collisional ionization equilibrium enriched to $Z=Z_{\odot}$ but consisting solely of C, O, Fe, and Si (in equal parts). Although this does not represent the specific yield of a PISN or a hypernova, we here only intend to give an approximate argument and defer a more precise treatment to future work.

Assuming the time-dependent properties of the simulation, we find that metal cooling is only briefly important after $10~\rm{Myr}$ (see Fig. 12), when the interior has adiabatically cooled to $10^{5}~\rm{K}$ and metal cooling is most efficient. At the same time, however, IC scattering becomes important and temperatures quickly fall to $10^{4}~\rm{K}$, where cooling rates drop by a few orders of magnitude. Due to the low densities in the interior, the onset of fine structure cooling below $10^{4}~\rm{K}$ also proves inefficient, and we conclude that even for initial metal yields as high as $y=0.1$, metal cooling is unimportant for the entire dynamical evolution of the SN remnant. We emphasize that the presence of metals will become crucially important once the enriched gas has recollapsed into a sufficiently massive halo later on, thereby reaching high densities again.

\subsection{Instabilities and Distribution of Metals}
Due to inefficient cooling, the evolution of the metal-enriched, interior bubble is governed by adiabatic expansion and preferentially propagates into the cavities created by the shock (see Figs. 8 and 14). Once the shock leaves the host halo and becomes highly anisotropic, the interior adopts the same behavior and expands into the voids surrounding the y-z plane in the shape of an hourglass, with a maximum extent similar to the final mass-weighted mean shock radius. This behavior is evident in Figure 14, where we plot the hydrogen number density and distribution of metal particles at various times after the SN explosion. Furthermore, since the radius of each metal particle is determined by the smoothing length, we find that at late times the interior becomes substantially mixed with the initial stellar ejecta.

When the shock finally stalls, the interior bubble is in pressure equilibrium with its surroundings, but it stays confined within the dense shell. To investigate the importance of RT instabilities in this configuration, we estimate the mixing length $\lambda_{\rm{rt}}$ for large density contrasts between two media according to \citep[e.g.,][]{mfr01}:
\begin{equation}
\lambda_{\rm{rt}}\simeq2\pi gt_{\rm{sh}}^{2}\mbox{\ ,}
\end{equation}
where $g$ is the gravitational acceleration of the host halo. Estimating the distance of the dense shell from the host halo with the final mass-weighted mean shock radius, we find that mixing between the dense shell and the interior bubble takes place on scales $\la 10~\rm{pc}$ in the course of a few times $10~\rm{Myr}$. In light of the substantial extent of the interior, we conclude that such mixing is generally inefficient and that much larger potential wells must be assembled to recollect and mix all components of the shocked gas. Specifically, turbulence arising in the virialization of the first galaxies could be an agent for this process \citep[e.g.,][]{wa07}.

On the other hand, mixing could be important in the y-z plane, where, due to the progression of structure formation, material falls in along filaments toward the metal-enriched gas near the host halo. Since this mechanism will only affect a limited volume, while most metals are expelled into the voids (see Fig. 14), we again conclude that a fundamental transition in star formation from Pop~III/Pop~II.5 to Pop~II requires the assembly of a DM halo with a sufficiently deep potential well.

To find the minimum mass necessary to recollect the hot and underdense post-shock gas residing at $T\sim 10^{3}~\rm{K}$ and $n_{\rm{H}}\sim 10^{-2.5}~\rm{cm}^{-3}$ (see \S3.4), we estimate by means of the cosmological Jeans criterion that a DM halo of at least $M_{\rm{vir}}\ga 10^{8}~M_{\odot}$ must be assembled. Assuming that in the course of its virialization the initial stellar ejecta mix with the entire swept-up mass, we find a final, average metallicity of $Z\simeq 10^{-2.5}Z_{\odot}$. Although this value is well above $Z_{\rm{crit}}$, we emphasize that the final topology of metal enrichment could be highly inhomogeneous, with pockets of highly enriched material on the one hand, but regions with a largely primordial composition on the other hand.

\section{Summary and Conclusions}
We have investigated the explosion of a $200~M_{\odot}$ PISN in the high-redshift universe by means of three-dimensional, cosmological simulations, taking into account all necessary chemistry and cooling. Using a ray-tracing algorithm to determine the size and structure of the H~{\sc ii} region around the progenitor star, we have followed the evolution of the SN remnant until it effectively dissolves into the IGM, and discussed its expansion and cooling properties in great physical detail. Specifically, we have found that a chronological sequence in its evolution, based on various physical mechanisms becoming dominant, allows the introduction of a simple analytic model summarizing its expansion properties. The SN remnant propagates for a Hubble time at $z\simeq 20$ to a final mass-weighted mean shock radius of $2.5~\rm{kpc}$, roughly half the size of the H~{\sc ii} region, and in this process sweeps up a total gas mass of $2.5\times 10^{5}~M_{\odot}$. We have found that its morphology becomes highly anisotropic due to encounters with filaments and neighboring minihalos in the y-z plane, but underdense voids perpendicular to the y-z plane. Based on the high explosion energy, the host halo is entirely evacuated, while in our case shock compression of neighboring minihalos partially offsets the delay in star formation due to negative feedback from photoionization heating. In contrast, we do not observe gravitational fragmentation triggered by efficient cooling behind the SN shock, which could in principle lead to secondary star formation. We have found that the metal-enriched, interior bubble expands adiabatically into the cavities created by the shock and preferentially propagates into the voids of the IGM with a maximum extent similar to the final mass-weighted mean shock radius. We have estimated that RT instabilities do not efficiently mix the dense shell with the interior, but that material falling in along filaments could mix with metal-enriched gas near the host halo. Finally, we have concluded that a DM halo of at least $M_{\rm{vir}}\ga 10^{8}~M_{\odot}$ must be assembled to recollect and mix all components of the shocked gas.

Based on the simulation, we find that a single PISN can enrich the local IGM to a substantial degree. If energetic SNe were indeed a common fate for the first stars, they might have deposited metals on large scales before massive galaxies formed and outflows were suppressed by their increasingly deep potential wells. Hints on ubiquitous metal enrichment have recently been found in the low column density Ly$\alpha$ forest \citep[e.g.,][]{sc96,songaila01,aguirre05} and in dwarf spheroidal satellites of the Milky Way \citep{helmi06}, where the presence of a ``bedrock'' metallicity was inferred. Based on these observations, various authors have argued that SNe occurring in the shallow potential wells of minihalos could substantially preenrich the universe \citep[e.g.,][]{daigne04,daigne06,ybh04,mc05,gb06}. Although the frequency of PISNe is debated \citep[e.g.,][]{ssf03,vt03,nop04,ro04}, we have found that the overall dynamics of the SN remnant and the distribution of metals are largely independent of the progenitor and are governed mainly by the explosion energy. Our simulations would therefore also approximately describe a hypernova explosion, where a rotating, massive star undergoes core collapse \citep[e.g.,][]{un02,tun07}. Such scenarios might better explain the peculiar yields found in extremely metal-poor stars in the Galactic halo \citep[e.g.,][]{christlieb02,frebel05}, assuming they formed out of gas enriched by a previous generation of stars \citep[e.g.,][]{un03,un05,tvs04,iwamoto05,karlsson06,tumlinson06}. However, in light of the tentative identification of SN~2006gy as a PISN \citep{smith07}, we find that the plausibility of the PISN scenario is enhanced and that it provides a viable possibility for the ultimate fate of a very massive, metal-free star.

When and where do the first Pop~II stars form? Based on the progression of structure formation, we expect that the first low-mass stars will form at the centers of the first galaxies, where primordial material streams in along filaments and mixes with the metal-enriched gas of the host halo. On the other hand, metals expelled into the voids are not so readily available, as the associated gas resides at too low densities and high temperatures. Because of this diversity, secondary star formation in the virialization of the first galaxies will likely be a highly complicated and multi-faceted process. To understand the relevant mechanisms in detail, one must perform numerical simulations that include metal cooling as well as an efficient model for their mixing. Furthermore, one must take into account the effects of multiple star-forming regions, including the expansion of additional H~{\sc ii} regions and SN remnants. In future work, we hope to shed some light on these issues by performing detailed numerical simulations incorporating all the necessary physics and evolving them to sufficiently late times. Such investigations are key in arriving at detailed predictions for the properties of the first galaxies to be observed with {\it JWST}.

\acknowledgements{We thank N. Yoshida for helpful comments which improved the content and layout of this work. We are grateful to the staff of the ACES Visualization Laboratory for extensive support in visualization and imaging software. T. H. G. would like to thank R. Banerjee and K. Vega for invaluable help in creating the images presented in this paper. All simulations were carried out at the Texas Advanced Computing Center (TACC).}

\bibliographystyle{apj}

\end{document}